\documentclass[%
reprint,mathtools,amsmath,amssymb,
 aps,pre,amsfonts
]{revtex4-2}
\usepackage{dcolumn}
\usepackage{bm}
\usepackage{xcolor}
\usepackage{mathrsfs}
\usepackage{braket}

\usepackage{eso-pic}
\usepackage{calc}
\usepackage{upgreek}
\usepackage{blindtext}
\usepackage{graphicx}%
\usepackage{float}

\begin{document}

\preprint{APS/123-QED}

\title{Multifractal dimensions for orthogonal-to-unitary crossover ensemble}

\author{Ayana Sarkar}
\email{ayanas1994@gmail.com}
\thanks{Present address: D\'{e}partement de physique and Institut Quantique, Université de Sherbrooke, Sherbrooke, QC, J1K 2R1, Canada}

\author{Ashutosh Dheer}%
\email{ad326@snu.edu.in}

\author{Santosh Kumar}%
\email{skumar.physics@gmail.com}
\affiliation{Department of Physics, Shiv Nadar Institution of Eminence (SNIoE), Gautam Buddha Nagar, Uttar Pradesh - 201314, India.}

\begin{abstract}
Multifractal analysis is a powerful approach for characterizing ergodic or localized nature of eigenstates in complex quantum systems. In this context, the eigenvectors of random matrices belonging to invariant ensembles naturally serve as models for ergodic states. However, it has been found that the finite-size versions of multifractal dimensions for these eigenvectors converge to unity logarithmically slowly with increase in the system size $N$. In fact, this strong finite-size effect is capable of distinguishing the ergodicity behavior of orthogonal and unitary invariant classes. Motivated by this observation, in this work, we provide semi-analytical expressions for the ensemble-averaged multifractal dimensions associated with eigenvectors in the orthogonal-to-unitary crossover ensemble. Additionally, we explore shifted and scaled variants of multifractal dimensions, which, in contrast to the multifractal dimensions themselves, yield distinct values in the orthogonal and unitary limits as $N\to\infty$ and therefore may serve as a convenient measure for studying the crossover. We substantiate our results using Monte Carlo simulations of the underlying crossover random matrix model. We then apply our results to analyze the multifractal dimensions in a quantum kicked rotor, a Sinai billiard system, and a correlated spin chain model in a random field. The orthogonal-to-unitary crossover in these systems is realized by tuning relevant system parameters, and we find that in the crossover regime, the observed finite-dimension multifractal dimensions can be captured very well with our results.
\end{abstract}

                          
\maketitle

\section{Introduction} 
\label{SecIntro}

Eigenvalues and eigenvectors play a pivotal role in quantum mechanics by providing essential information about the energy states and observable properties of quantum systems, enabling reliable predictions and analyses of their behavior. Unfortunately, except for a few simple physical systems, it is not feasible to compute the eigenfunctions in quantum many-body problems analytically. As a result, diverse statistical and numerical methods become indispensable. In this context, techniques from random matrix theory (RMT) are of particular importance. Statistical exploration of eigenvectors dates back to studies pertaining to the transition amplitudes and fluctuations of nuclear reaction widths in complex nuclei~\cite{PT1956,Wigner1967,BFFMPW1981,Mehta2004}. Over time, the quantification of chaotic or regular and ergodic or non-ergodic nature of eigenvectors has emerged as a central theme in the study of quantum chaotic and complex many-body systems, such as quantum billiards, quantum maps, and quantum spin chains, among others~\cite{BHK2019,Berry1977,SG1984,MK1988,AS1991,AS1993,HZ1990,LR1994,SE1996,Prosen1997,Backer2003,Izr1987,Non1997,NV1998,Non2013,KMH1988}. The behaviour of eigenvectors also sheds light on crucial phenomena like thermalization and many body localization~\cite{Deu1991,Sred1994,BAA2006,BN2013,NH2015,AP2017,AABS2019}. Within RMT, the results for the distribution of a generic eigenvector component for the canonical Gaussian and circular ensembles are known, for example, from Refs.~\cite{BFFMPW1981,KMH1988,HZ1990}. The crossover random matrix ensembles have remained more or less unexplored in this regard, except for the Gaussian orthogonal to unitary crossover case, for which the distribution of a generic eigenvector component has been derived by Sommers and Iida in Ref.~\cite{SI1994}. Among the various avenues for exploring the behavior of eigenvectors, along with the full distribution, inverse participation ratio (IPR) or equivalently the number of principal components, information entropy and multifractal dimensions, hold a significant position.

The notion of multifractality is derived from the much discussed fractal dimension in nonlinear dynamics and classical physics. According to Mandelbrot, a fractal dimension serves as an index to characterize fractal patterns or sets by quantifying their complexity through the ratio of detail change to scale change~\cite{Mand1983}. Some of the earliest work on the mathematics of fractals was brought forth by Richardson and Mandelbrot in the context of the \emph{coastline paradox}~\cite{Vul2014,Rich1961,Mand1967,Mand1983,Rich1993}. The notion of multifractal dimensions is used for systems where a single exponent is insufficient to capture their structure or dynamics. Multifractal behaviour is common in nature and is crucial in the quantitative understanding of irregular constructs appearing in varied areas of classical physics~\cite{Sree1991,SM1988,JXZS2019}.
In quantum physics, especially in the context of many-body systems, multilfractal dimensions provide a succinct description of eigenvector moments and serve as simple indicators of wave function spreading at different scales. Multifractility-related measures have already been extensively utilized in problems related to Anderson transitions (including both metal-insulator transitions and quantum-Hall-type transitions)~\cite{SG1991,H1995,ME2000,EM2008}, thermalization and many-body localization~\cite{T2020,PL2021,OI2021}. Appearance of multifractility has been discovered in quantum spin chain systems~\cite{AB2012,AB2014,Mon2016,Voliotis2019,MAL2019,LKL2020,SSH2021}, Bose-Hubbard model~\cite{LBR2019,PCRB2021}, quantum maps~\cite{MGG2008,MGGG2010,BGGG2019}, quantum start graphs, quantum kicked top model~\cite{WR2021}, and various random matrix models~\cite{HP2000,BG2011,BG2012,KKCA2015,Bog2020,BG2021}.  

In Ref.~\cite{BHK2019}, B\"acker, Haque, and Khaymovich have considered circular orthogonal ensemble (COE) and circular unitary ensemble (CUE) of random matrices, and analytically obtained the finite-size dependence of the mean behavior of the multifractal dimensions via averaged eigenvector moments, which provides a lower bound to the typical (logarithmic) averages. These analytical results confirm that as the dimension $N$ of the matrices tend to infinity, these mean multifractal dimensions tend to the value 1, thereby asserting the ergodic nature of the eigenvectors associated with invariant orthogonal and unitary ensembles. However, the convergence of these mean multifractal dimensions to unity is logarithmically slow. As a consequence even for large but finite $N$, very strong finite-size effects are noticed. Interestingly, these finite size effects are capable of distinguishing the orthogonal and unitary symmetry classes. This is the primary motivation behind our present work, where we try to exploit this finite-dimension effect to quantify the ergodicity of quantum chaotic or many-body systems lying in the crossover regime, for instance due to the presence of a weak magnetic field. Additionally, we propose shifted and scaled variant of the multifractal dimensions which lead to distinct result in the $N\to\infty$ limit and therefore may serve as more convenient choices to discern the orthogonal and unitary classes and to examine the crossover.

In this paper, we present semi-analytical expressions for the finite-dimensional ensemble-averaged multifractal dimensions and the associated shifted and scaled quantities in the orthogonal to unitary crossover ensemble. We validate our findings by using Monte Carlo simulations of the associated Gaussian orthogonal ensemble (GOE) to Gaussian unitary ensemble (GUE) crossover random matrix model. Subsequently, we apply our results to investigate the multifractal dimensions in the quantum kicked rotor model, a quarter Sinai billiard, and a correlated spin chain system in a random magnetic field. Our findings demonstrate that the random matrix theory (RMT) results effectively capture the ergodic behavior of these systems in the crossover regime. The structure of the rest of this paper is as follows. In Sec. \ref{MFdef}, we provide a concise overview of the mathematical formulation pertaining to multifractal dimensions and their ensemble-averaged counterparts derived from the eigenvector moments. Next, in Sec. \ref{MFoeue}, we compile the formulae for the distribution of generic eigenvector components in Gaussian or circular ensembles, along with the resulting expressions for ensemble-averaged finite-dimensional multifractal dimensions derived in Ref. \cite{BHK2019}. Subsequently, in Sec. \ref{MFouce}, we employ the distribution of a generic eigenvector component in the orthogonal-to-unitary crossover random matrix model to derive semianalytical results for the ensemble-averaged multifractal dimensions. We validate these results for the eigenvector distribution and multifractal dimensions through simulations of the crossover random matrix model, as discussed in Sec. \ref{numres-RMT}. Following that, we apply our results to examine multifractal dimensions in quantum chaotic, and many-body systems in Sec. \ref{QCMBS}. Finally, we provide a summary of our work and discuss potential future directions in Sec. \ref{summ}.

\begin{figure*}[!ht]
  \centering
\includegraphics[width=0.8\linewidth]{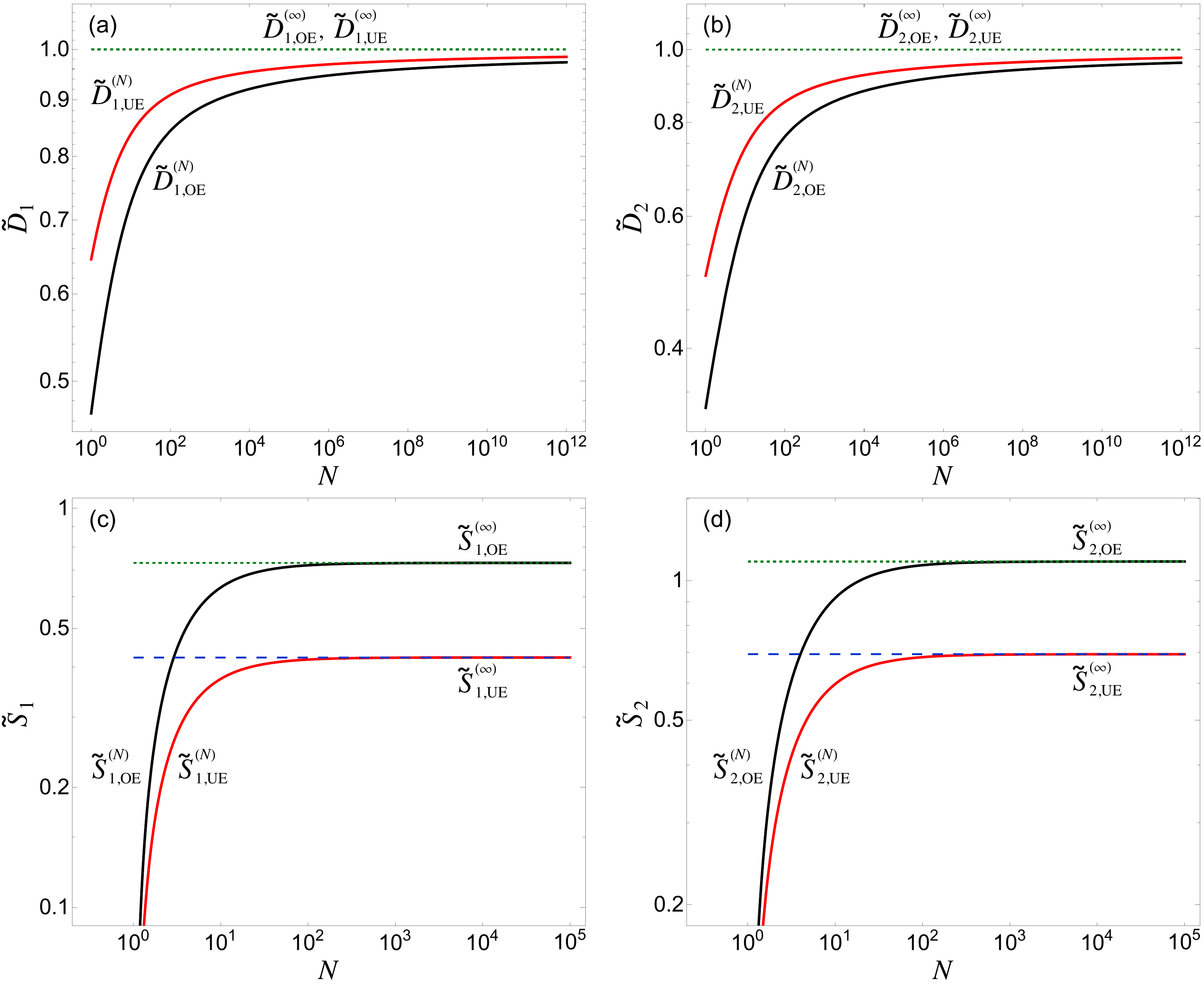}  \caption{Log-log plot of fractal dimensions (a) $\widetilde{D}_{1}^{(N)}$, (b) $\widetilde{D}_{2}^{(N)}$ and the associated shifted and scaled quantities (c) $\widetilde{S}_{1}^{(N)}$, (d) $\widetilde{S}_{2}^{(N)}$, with respect to the system size $N$, for both orthogonal and unitary ensembles. Very slow convergence of $\widetilde{D}_{1}^{(N)}$ and $\widetilde{D}_{2}^{(N)}$ towards the value of 1 is seen as $N$ is increased and it is possible to discern the two universality classes for finite $N$. On the other hand, the quantities $\widetilde{S}_{1}^{(N)}$ and $\widetilde{S}_{2}^{(N)}$ converge to distinct values at a relatively faster rate, making them useful alternatives for distinguishing between the orthogonal and unitary symmetry classes and for studying the crossover between them.}
  \label{fig1}
  \end{figure*}

\section{Multifractal dimensions: Preliminaries}
\label{MFdef}
Let us consider a given quantum state $|\Psi_{j}\rangle$, expanded in some finite $N$-dimensional orthonormal basis $\{|\psi_{i}\rangle\}$ such that  $|\Psi_{j}\rangle =\sum\limits_{i=1}^{N}c_{i}^{(j)} |\psi_{i}\rangle$. The associated moments are defined as 
\begin{equation}
I_{q} (j) = \sum\limits_{i = 1}^{N} |c_{i}^{(j)}|^{2q}.
\end{equation}  
The finite-$N$ fractal dimensions or generalized information entropies for the considered state can be expressed in terms of the above moments as, 
\begin{equation}
D_{q}^{(N)}(j) = -\frac{1}{(q-1)\ln N}\ln I_{q}(j).
\end{equation}
For finite $N$, the values of $D_{q}^{(N)}$ lie in the interval $[0,1]$ and are monotonically decreasing with increasing $q$ for $q\geq 0$. 
The first moment $I_1(j)$ is related to the first fractal dimension $D_{1}^{(N)}$ which, besides the factor $1/\ln N$, is identified the Shannon information entropy or the von Neumann entropy,
\begin{equation}
\label{D1eq}
 D_{1}^{(N)}(j) = -\frac{1}{\ln N}\sum\limits_{i = 1}^{N} |c_{i}^{(j)}|^{2}\ln  |c_{i}^{(j)}|^{2}.
\end{equation}
Similarly, the second moment $I_2(j)$ gives the second $N$-dependent fractal dimension $D_{2}^{(N)}$ which is related to the logarithm of the well known inverse-participation ratio,
\begin{equation}
\label{D2eq}
  D_{2}^{(N)}(j) = -\frac{1}{\ln N}\ln\left(\sum\limits_{i = 1}^{N}(c_{i}^{(j)})^{4}\right).
\end{equation}
It can be used to study the degree of localization or delocalization of states in Hilbert space.

Conventionally, what one calls multifractal dimensions are the $N\to\infty$ limit of the above $N$-dependent expressions, i.e., multifractal dimensions are defined as $D_{q}^{(\infty)}=\lim\limits_{N\rightarrow \infty}D_{q}^{(N)}$~\cite{BHK2019,ME2000}. If $D_{q}^{(\infty)}$ exhibit a nontrivial dependence on $q\, (>  0)$ in a nontrivial way, the states are multifractal, whereas for constant $D_{q}^{(\infty)}<1$ the states are simply fractal. The extreme cases of $D_{q}^{(\infty)}=0$ ($q\ne 0$) and $D_{q}^{(\infty)}=1$ correspond to the cases of localized and ergodic states, respectively. Multifractal dimensions are usually studied for positive integer $q$ values, however researchers have also examined real $q$ values including, negative ones, to obtain a fuller picture of behavior of the eigenstates~\cite{BG2011,BG2012,MGG2008,MGGG2010}.

To obtain the RMT predictions for the multifractal dimensions, the average of the above results over an adequate ensemble of random matrices is considered. This procedure of \emph{ensemble averaging} becomes indispensable when one is dealing with finite-dimension not-so-large-$N$ cases. However, as argued in Ref.~\cite{BHK2019}, the presence of the logarithm of the moments in the expression of $D_{q}^{(N)}$, makes this averaging nontrivial to perform analytically. Therefore, one considers ensemble-averaged moments $\widetilde{I}_{q}^{(N)} = \langle I_{q}^{(N)} (j) \rangle$ and take the logarithm afterwards. This gives,
\begin{align}
\label{Dqeq}
\widetilde{D}_{q}^{(N)} = -\frac{1}{(q-1)\ln N} \ln \widetilde{I}_{q}^{(N)},
\end{align} 
which due to the Jensen's inequality, provides a lower bound to $D_{q}^{(N)}$. In this work, among all $D_{q}^{(N)}$'s, we particularly focus on $\widetilde{D}_{1}^{(N)}$ and $\widetilde{D}_{2}^{(N)}$.

\section{Multifractal dimensions for orthogonal and unitary ensembles}
\label{MFoeue}

In the context of quantum chaos, ergodic states are expected to follow RMT-based statistics. For such systems, the eigenfunctions are spread uniformly over all components with respect to any basis. Previously, the distribution of one eigenvector component $c_{i}^{(j)}$ has been investigated~\cite{BFFMPW1981,KMH1988,HZ1990},
in the form of squared-modulus $x = N|c_{i}^{(j)}|^{2}$ for the Gaussian (or circular) ensemble. One has,
\begin{equation}
\label{pcoeex}
P_\mathrm{OE}(x) = \sqrt{N}\frac{\Gamma\left(N/2\right)}{\Gamma\left((N-1)/2\right)} \frac{\left(1-\frac{x}{N}\right)^{\frac{N-3}{2}}}{\sqrt{\pi x}},
\end{equation}
\begin{equation}
\label{pcueex}
P_\mathrm{UE}(x) = (N-1)\left(1-\frac{x}{N}\right)^{N-2}.
\end{equation}
where OE and UE denote the orthogonal ($\beta = 1$) and unitary ($\beta = 2$) symmetry classes, respectively. In the limit of $N \rightarrow \infty$, the above distributions take the following form,
\begin{equation}
\label{evGOE}
\hat{P}_\mathrm{OE}(x) = \frac{1}{\sqrt{2\pi x}} e^{-x/2},
\end{equation}
\begin{equation}
\label{evGUE}
\hat{P}_\mathrm{UE}(x) = e^{-x}.
\end{equation}
These two expressions are special cases of the gamma-distribution and also identified as the famous Porter-Thomas distribution~\cite{PT1956}. 

In Ref.~\cite{BHK2019}, Eqs.~\eqref{pcoeex} and \eqref{pcueex} have been used to derive the corresponding moments of the eigenvectors and multifractal dimensions for orthogonal and unitary ensembles. The ensemble averaged $N$-dependent multifractal dimensions are given by,
\begin{align}
\label{fractalCOEq}
\nonumber
\widetilde{D}_{q,\mathrm{OE}}^{(N)} &= -\frac{1}{(q-1)\ln N}\ln\bigg[ \frac{N \Gamma(N/2)\Gamma(q+1/2)}{\sqrt{\pi}\Gamma(q+N/2)}\bigg] \\
&\simeq 1-\frac{1}{(q-1)\ln N}\ln\bigg[\frac{\Gamma(q+1/2)2^{q}}{\sqrt{\pi}}\bigg],
\end{align}
\begin{align}
\label{fractalCUEq}
\nonumber
\widetilde{D}_{q,\mathrm{UE}}^{(N)} &= \frac{1}{(q-1)\ln N} \ln \bigg[\frac{q! N!}{(N-1+q)!}\bigg] \\
&\simeq1-\frac{\ln(q!)}{(q-1)\ln N},
\end{align}
for $q\geq2$. Here, the expressions in the second line of both equations are the corresponding large-$N$ results. The analytical expressions for $\widetilde{D}_{1,\mathrm{OE}}^{(N)}$ and $\widetilde{D}_{1,\mathrm{UE}}^{(N)}$ are obtained by considering the limit $q \rightarrow 1$,
\begin{align}
\label{fractalCOE1}
\nonumber
\widetilde{D}_{1,{\mathrm{OE}}}^{(N)} &= \frac{\uppsi[(N+2)/2]-\uppsi(3/2)}{\ln N}\\
&\simeq 1-\frac{\ln 2+\uppsi(3/2)}{\ln N} ,
\end{align}
\begin{align}
\label{fractalCUE1}
\nonumber
\widetilde{D}_{1,\mathrm{UE}}^{(N)} &= \frac{-1+\upgamma+\uppsi(N+1)}{\ln N}\\
& \simeq 1-\frac{1-\upgamma}{\ln N}.
\end{align}
Here, $\uppsi(x)$ is the digamma function and $\upgamma\approx 0.5772$ is the Euler--Mascheroni constant. It should be noted that the above results hold for both Gaussian and circular ensembles. 

Along with the fractal dimensions $D_q^{(N)}$ and its ensemble averaged variant $\widetilde{D}_q^{(N)}$ We also consider the associated shifted and scaled quantities 
\begin{align}
S_q^{(N)}=\ln N\cdot(1-D_q^{(N)}),\\
\widetilde{S}_q^{(N)}=\ln N\cdot(1-\widetilde{D}_q^{(N)}).
\end{align}
 These quantities lead to distinct limits for the OE and UE cases when $N\to\infty$, unlike the fractal dimensions which in both cases tend to 1. Specifically, from Eqs.~\eqref{fractalCOEq}-\eqref{fractalCUE1}, we have for $q\ge2$,
\begin{equation}
\label{SqOE}
\widetilde{S}_{q,\mathrm{OE}}^{(\infty)}=\frac{1}{q-1}\ln\bigg[\frac{\Gamma(q+1/2)2^{q}}{\sqrt{\pi}}\bigg],
\end{equation}
\begin{equation}
\label{SqUE}
\widetilde{S}_{q,\mathrm{UE}}^{(\infty)}=\frac{\ln(q!)}{q-1},
\end{equation}
and for $q=1$,
\begin{equation}
\label{S1OE}
\widetilde{S}_{1,\mathrm{OE}}^{(\infty)}=\ln 2+\uppsi(3/2),
\end{equation}
\begin{equation}
\label{S1UE}
\widetilde{S}_{1,\mathrm{UE}}^{(\infty)}=1-\upgamma.
\end{equation}
The test of ergodicity of a system in terms of these quantities would be matching the empirical values for various $q$ with the above RMT based values.

In Fig.~\ref{fig1} we plot the fractal dimensions $\widetilde{D}_{1}^{(N)}, \widetilde{D}_{2}^{(N)}$ and also $\widetilde{S}_{1}^{(N)}, \widetilde{S}_{2}^{(N)}$ as functions of $N$ for both OE and UE classes. As asserted earlier, we can see logarithmically slow convergence of the fractal dimensions towards unity with increasing $N$. However, it is possible to distinguish between the two symmetry classes due to the finiteness of $N$, even for its extremely large values. The crossover regime is also discernible due to this finite size effect, as demonstrated ahead. The shifted and scaled quantities $\widetilde{S}_{1}^{(N)}$ and $\widetilde{S}_{2}^{(N)}$ provide additional benefit that they approach distinct values for the OE and UE classes as $N\to\infty$. Moreover, their convergence towards the corresponding $\widetilde{S}_{1}^{(\infty)}$ and $\widetilde{S}_{2}^{(\infty)}$ values is fast. Therefore, these quantities serve as convenient measures to examine the symmetry crossover.

\section{Results for orthogonal-to-unitary crossover ensemble}
\label{MFouce}

Besides the orthogonal and unitary cases, analytical result for the distribution of one eigenvector component is known for the orthogonal-to-unitary crossover ensemble due to Sommers and Iida~\cite{SI1994}. The crossover is modeled for the Gaussian ensemble using the Pandey-Mehta Hamiltonian~\cite{PM1983,MP1983},
\begin{align}
\label{PM1}
H = S + i \alpha A.
\end{align}
Here, $S$ and $A$ are $N$-dimensional real-symmetric and real-antisymmetric, statistically independent random matrices, respectively. The matrix elements of $S$ and $A$ are zero-mean Gaussian variables specified by variances $\langle (S_{m \leq n})^{2} \rangle = (1 + \delta_{m,n}) v^{2}$ and $\langle (A_{m< n})^{2} \rangle  = v^{2}$, where, as before, $\langle~ \rangle$ represents the ensemble average. The crossover is governed by the parameter $\alpha$ which gives GOE ($\beta$=1) for $\alpha = 0$ and GUE $(\beta=2)$ for $\alpha = 1$. The above matrix model is statistically equivalent to
\begin{align}
\label{PM2}
H = \sqrt{1-\alpha^{2}} \,H_{1} + \alpha H_{2}
\end{align}
where $H_{1}$ and $H_{2}$ are random matrices belonging to GOE (real-symmetric) and GUE (complex hermitian), respectively. The diagonal elements of both $H_{1}$ and $H_{2}$ are zero-mean Gaussians with variance $2v^2$, whereas their off-diagonal elements (both real and imaginary parts in the case of $H_2$) are zero-mean Gaussians having variance $v^2$. It should be noted that any choice of $v^2>0$ would not matter for the statistics of normalized eigenvectors, or if one wants to explore the local spectral fluctuations of eigenvalues. However, the eigenvalue density does depend on it, and is given by the Wigner semi-circle for large $N$, viz.,
\begin{equation}
\label{semicirc}
p(E)=\frac{2}{\pi R^2}\sqrt{R^2-E^2};~~ R=\sqrt{4Nv^2(1+\alpha^2)}.
\end{equation} 

\begin{figure}[!t]
  \centering
\includegraphics[width=0.9\linewidth]{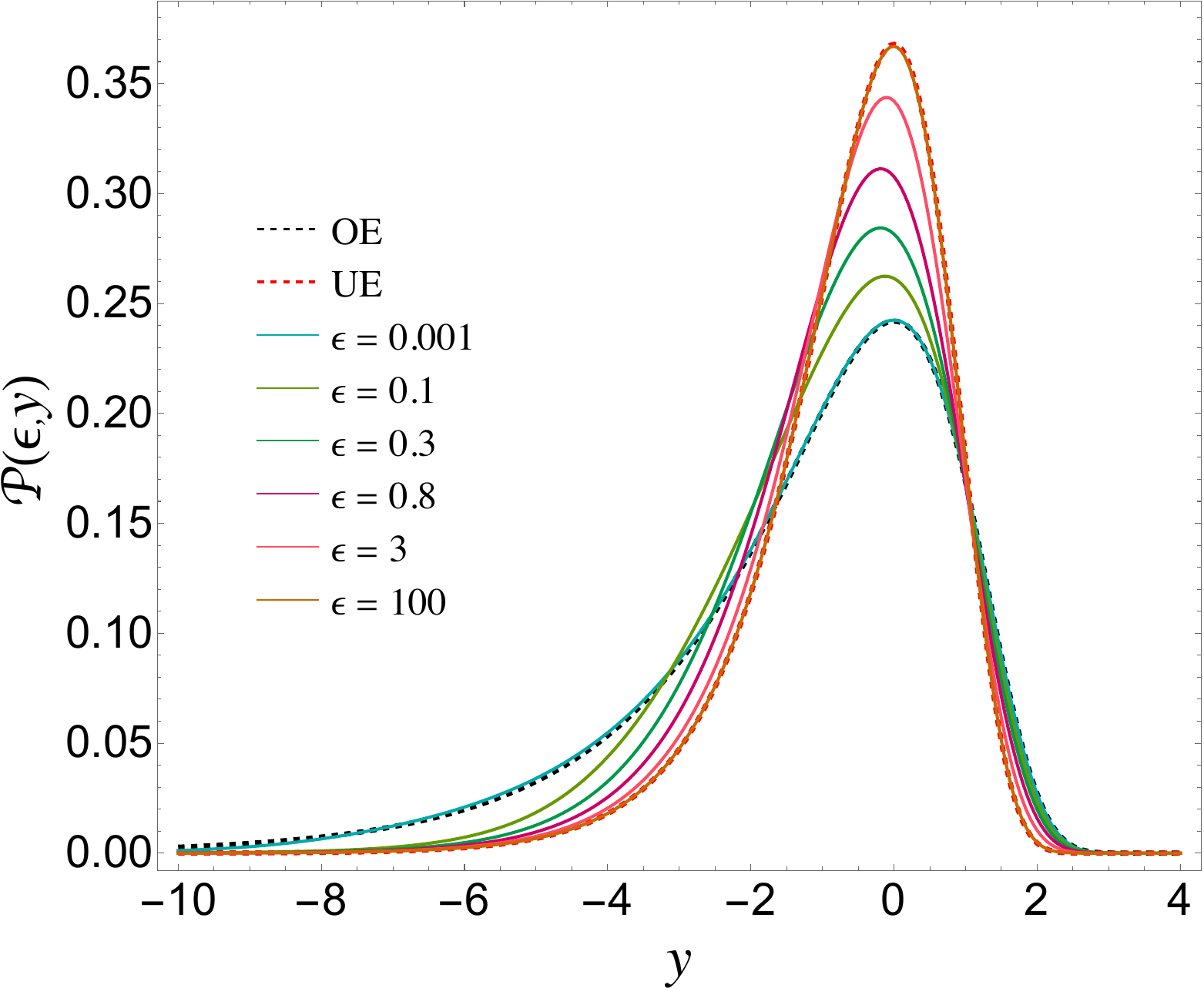}  \caption{Distribution of a generic eigenvector component in orthogonal-to-unitary crossover ensemble (solid lines) in terms of the variable $y=\ln x$, calculated using Eq.~\eqref{sommres}. The crossover is realized by tuning the rescaled parameter $\epsilon$. The extreme cases of OE and UE are shown using dashed black and red lines, respectively, and are based on Eqs.~\eqref{pcoeex} and \eqref{pcueex}.}
  \label{EvecCrossover}
  \end{figure}

The probability density function (PDF) of one rescaled component $x = N |c_{i}^{(j)}|^{2}$ of the eigenvector has been calculated in Ref.~\cite{SI1994} as a two-fold integral in the $N \rightarrow \infty$ limit,
\begin{align}
\label{evec-som-for}
P(\epsilon, x) = \frac{\epsilon e^{\epsilon}}{\pi} \int\limits_{0}^{\pi} d \phi  \int\limits_{0}^{\pi} d\theta \frac{\exp\left(-\frac{\epsilon + x(1-\cos \phi)}{\sin^{2}\phi \sin^{2}\theta }\right)}{\sin^{4}\theta \sin^{3}\phi},
\end{align}
where $\epsilon = \alpha^{2}N$ is the rescaled crossover parameter. It turns out that the $\theta$-integral can be solved in an elementary form using the result
$$\int\limits_{0}^{\pi} d\theta \frac{\exp(-a/\sin^{2}\theta)}{\sin^{4}\theta} = \frac{\sqrt{\pi}\,e^{-a}}{2 a^{3/2}} (2a + 1); ~\text{Re}(a)>0.$$
For using the above result in Eq.~\eqref{evec-som-for} we identify $
a = \frac{\epsilon + x(1-\cos\phi)}{\sin^{2}\phi} = [\epsilon + 2 x \sin^{2}(\phi/2)]\csc^{2}\phi$,
and therefore we can write $P(\epsilon,x)$ in terms of a single integral, viz.,
\begin{align}
\label{sommres}
P(\epsilon,x) &= \frac{\epsilon e^{\epsilon}}{2\sqrt{\pi}} \int\limits_{0}^{\pi} d\phi \frac{\exp\left[-(\epsilon + 2 x \sin^{2}(\phi/2))\csc^{2}\phi\right]}{\left[\epsilon + 2 x \sin^{2}(\phi/2)\right]^{3/2}} \nonumber\\
  &~~~~\times\left[2(\epsilon + 2 x\sin^{2}(\phi/2)) \csc^{2}\phi + 1\right].
\end{align}
The results for OE ($\beta = 1$) and UE ($\beta = 2$) cases given in Eqs.~\eqref{evGOE} and \eqref{evGUE} are included in this crossover result in the limits $\epsilon \rightarrow 0$ and $\epsilon \rightarrow \infty$, respectively. It is worth mentioning at this point that an ansatz for the eigenvector intermediate-statistics was proposed by \.Zyczkowski and Lenz in Ref.~\cite{ZL1991}, however, it does not fully agree with the above exact result of Sommers and Iida. In Fig.~\ref{EvecCrossover}, we show the above crossover result for various $\epsilon$ values using the variable $y=\ln x$, for which the PDF is given by $\mathcal{P}(\epsilon,y)=e^yP(\epsilon,e^y)$. The extreme limits of GOE and GUE are also shown in this plot. 

\begin{figure*}[!tbp]
  \centering
\includegraphics[width=0.8\textwidth]{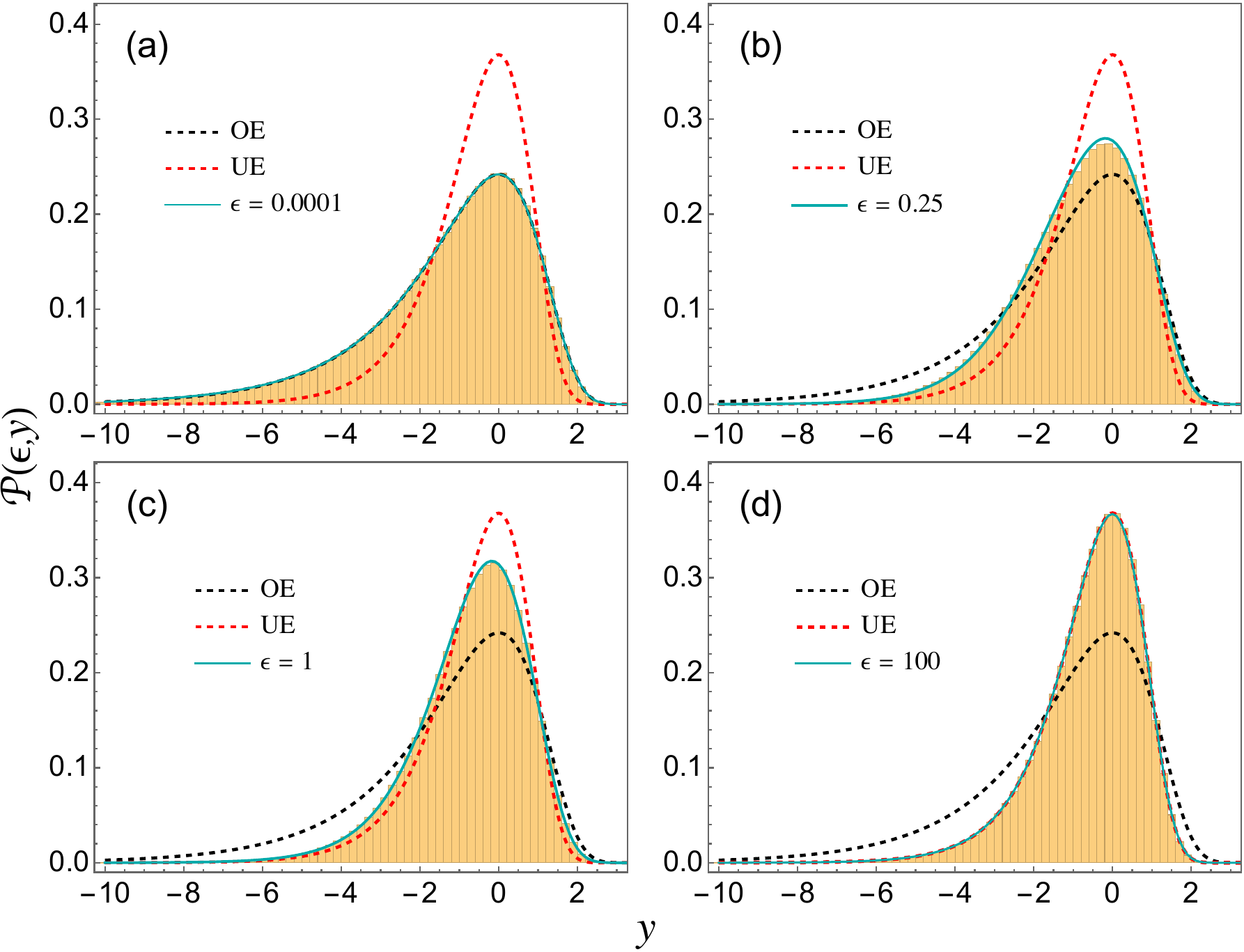}
  \caption{Distribution of a generic eigenvector component in orthogonal-to-unitary crossover: Comparison between numerical simulations (histograms) and analytical predictions (solid green lines). An ensemble comprising 200 matrices of dimension $N = 1000$ for various $\alpha$ values has been used for the simulation of the Pandey-Mehta matrix model, Eq.~\eqref{PM2}. The fits based on Eq.~\eqref{sommres} with $\epsilon=\alpha^2 N$ are seen to agree very well with the simulation results. The extreme cases of OE and UE, based on the analytical results in Eq.~\eqref{pcoeex} and \eqref{pcueex}, have been also shown using dashed black and red lines, respectively.}
  \label{GOE-GUE-MM-eigvec}
  \end{figure*}
Based on the above expression of probability density function of $x$, the corresponding moments can be calculated as,
\begin{align}
\label{moments}
\langle x^{q} \rangle &= \int\limits_{0}^{\infty} dx\, x^{q} P(\epsilon, x)  \nonumber \\
 &=  \frac{\Gamma(q+1)\epsilon^{q+1/2}}{2^{q+2}\sqrt{\pi}} \int\limits_{0}^{\pi} d\phi e^{-\epsilon \cot^{2}\phi} \csc^{2q +2} (\phi/2)\nonumber\\ 
 &~~ \times  [U(q+1,q+1/2, \epsilon \csc^{2}\phi)  \nonumber\\
 &~~~ + (2 \epsilon \csc^{2}\phi) U(q+1,q+3/2, \epsilon \csc^{2}\phi)],
\end{align}
where $U(a,b,z)$ is the confluent Hypergeometric function (Tricomi's function). 
The ensemble averaged moments can then be obtained as,
\begin{align}
\widetilde{I}_{q} (\epsilon) = N^{-(q-1)} \left \langle \left[N (c_{i}^{(j)})^{2}\right]^{q} \right \rangle = N^{-(q-1)} \langle x^{q}\rangle.
\end{align}
The variable $x$ is scaled by the dimension $N$ with the limit $N \rightarrow \infty$ already considered, therefore the explicit $N$-dependence is not retained in the expression for $\langle x^{q} \rangle$. However, based on the results of Ref.~\cite{BHK2019} for the orthogonal and unitary cases, it is anticipated that the dominating $N$-dependence in large $ N$-limit is contained in the pre-factor $N^{-(q-1)}$. Correspondingly, the fractal dimension is given using Eq.~\eqref{Dqeq} as,
\begin{align}
\label{Dqtilcr}
\widetilde{D}_{q}^{(N)} =1 -\frac{ \ln \langle x^{q} \rangle }{(q-1) \ln N}.
\end{align}
We note that the above expression cannot be directly used for the $q = 1$ case and instead a limiting procedure must be invoked. In fact, we can proceed from Eq.~\eqref{D1eq} itself and obtain the following result for $\widetilde{D}_{1}^{(N)}$,
\begin{align}
\label{D1tilav}
\widetilde{D}_{1}^{(N)} &= -\frac{N}{\ln N}\bigg\langle \frac{x}{N} \ln \frac{x}{N}\bigg\rangle = -\frac{1}{\ln N} \langle  x\ln x - x\ln N\rangle \nonumber \\
  &= -\frac{1}{\ln N} \langle x \ln x \rangle + \langle x \rangle = -\frac{1}{\ln N} \langle x \ln x \rangle + 1.
\end{align}
Therefore, using Eq.~\eqref{sommres}, we have
\begin{align}
\label{D1til}
\widetilde{D}_{1}^{(N)} = 1 - \frac{1}{\ln N}\int\limits_{0}^{\infty} dx\,x \ln x \widetilde{P}(\epsilon, x).
\end{align}
We note that the average $\langle x \ln x \rangle$ may be obtained from the moment as $\frac{\partial }{\partial b}\langle x^{b+1}\rangle|_{b=0}$. However, it is much more convenient to perform the integral in Eq.~\eqref{D1til} numerically. From Eqs.~\eqref{Dqtilcr} and~\eqref{D1til}, it is clear that for $N\to\infty$, $\widetilde{D}_{q}^{(N)}\to 1$, similar to the limiting orthogonal and unitary cases. However, for finite $N$ even when it is large, the deviation from the value of $1$ can be ascertained and therefore these results can be used for quantifying the extent of ergodicity in systems belonging to crossover regime. Here also, we define the shifted and scaled variants of fractal dimension for $N\to\infty$, viz.
\begin{equation}
\label{Sqinf}
\widetilde{S}_q^{(\infty)}=\frac{ \ln \langle x^{q} \rangle }{(q-1)}; ~~~q\ne1,
\end{equation}
and
\begin{equation}
\label{S1inf}
\widetilde{S}_1^{(\infty)}=\langle x \ln x \rangle.
\end{equation}
Unlike the fractal dimensions themselves as $N\to\infty$, these quantities lead to distinct values when considering $\epsilon\to0$ and $\epsilon\to 1$ limits, as in Eqs.~\eqref{SqOE}--\eqref{S1UE}.

The above crossover results hold for Gaussian ensemble. The crossover in the case of circular ensemble is describable in terms of Dyson's Brownian motion formalism~\cite{PS1991,FP1995}, however, it is complicated compared to that for the Gaussian case. Furthermore, there cannot be an additive formulation for the COE-to-CUE crossover such as Eq.~\eqref{PM1} or~\eqref{PM2}, and no similar description of eigenvectors in the crossover regime is available to the best of our knowledge. In Ref.~\cite{FP1995}, Frahm and Pichard have examined the relationship of the unitary-Brownian motion model with Pandey-Mehta model using Mahaux-Weidenm\"uller formula~\cite{MW1969} and obtained the connection of Brownian motion time of the COE-to-CUE crossover with the Pandey-Mehta crossover parameter. See also Ref.~\cite{ZK1996} for another kind of COE-to-CUE interpolating ensemble. For our purpose, considering the fact that in the extreme limits of OE and UE, the eigenvector statistics is identical for Gaussian and circular ensembles, we may use Eq.~\eqref{sommres} as a phenomenological formula for systems exhibiting COE-to-CUE crossover. The fitting-parameter $\epsilon$ in this case may be related to the system (physical) parameter and tuned to fit the empirical eigenvector distribution in the crossover regime. Such an example is considered in Sec.~\ref{numre-QKR} using the paradigmatic quantum kicked rotor model. In fact, for this system, validity of Eq.~\eqref{sommres} has been argued by Shukla for strongly chaotic case in great detail in Ref.~\cite{Shukla1996}.
 
\section{Comparison with RMT simulations}
\label{numres-RMT}

In this section, we compare the analytical predictions with the simulation results obtained from the crossover random matrix model, as given in Eqs.~\eqref{PM1} and~\eqref{PM2}. Tuning the parameter $\alpha$ in the matrix model leads to a transition from the orthogonal to the unitary symmetry class and the corresponding statistics may be verified by setting $\epsilon =  \alpha^{2} N$ in the analytical results by considering large-$N$ values.
 
At first we inspect the eigenvector distribution by simulating the GOE-to-GUE crossover matrix model using an ensemble comprising 200 matrices of dimension $N = 1000$ and using various values of the crossover parameter $\alpha$. For our simulation, we have considered the choice $v^2=[4N(1+\alpha^2)]^{-1}$ so that most of eigenvalues fall in the domain $[-1,1]$, except for some outlier edge-eigenvalues. The corresponding eigenvector distribution results have been shown in Fig.~\ref{GOE-GUE-MM-eigvec} using histograms. These are based on combining the datasets of all eigenvectors together. The green solid curves are fits based on Eq.~\eqref{sommres} with $\epsilon=\alpha^2 N$ and agree well with the crossover-regime simulation results. The extreme cases of GOE and GUE, based on Eqs.~\eqref{evGOE} and \eqref{evGUE}, have been also plotted with dashed black and red lines, respectively. 

\begin{figure*}[!tbp]
  \centering
\includegraphics[width=0.9\linewidth]{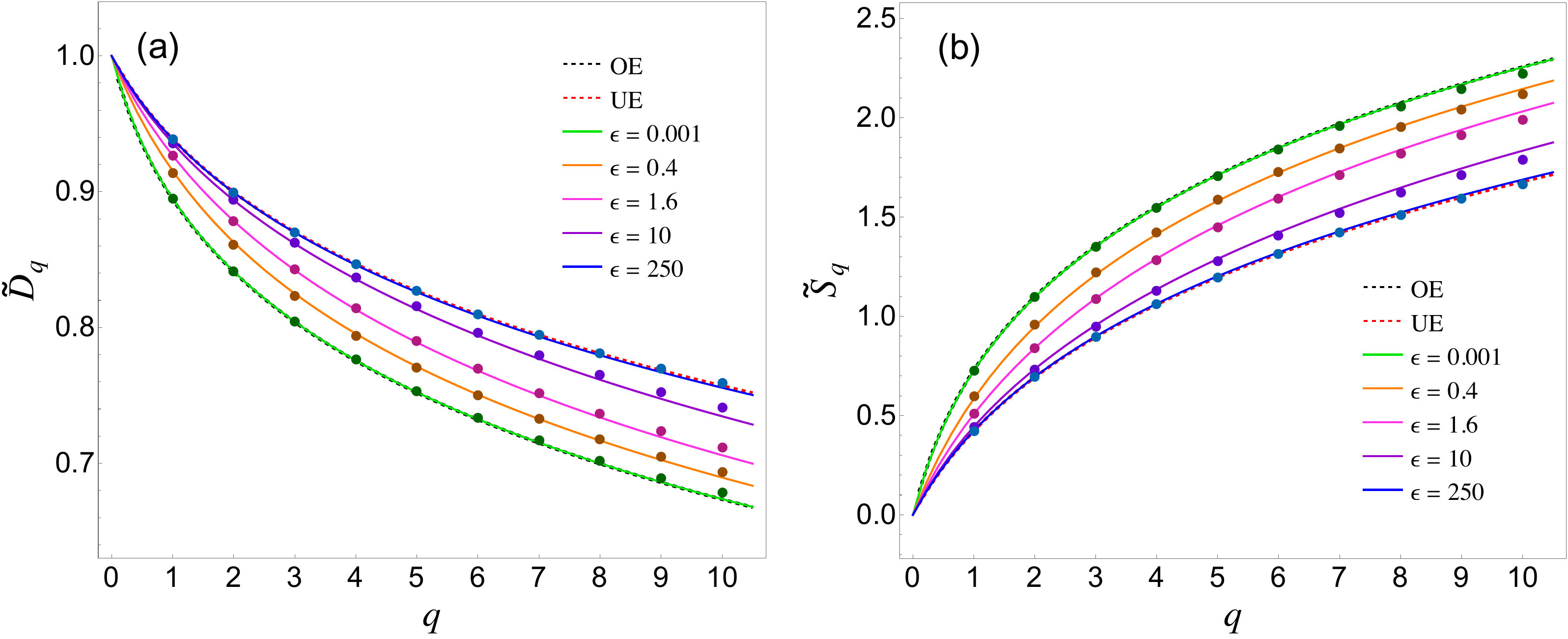}
  \caption{Plots of (a) multifractal dimensions $\widetilde{D}_{q}$ and the associated shifted and scaled quantities $\widetilde{S}_{q}$ for varying $q$ in the orthogonal to unitary crossover ensemble: Comparison between numerical simulations (disks) and analytical predictions (solid lines). The numerical results are based on 200 matrices of dimension $N = 1000$, as in Fig.~\ref{GOE-GUE-MM-eigvec}. The extreme cases of OE and UE, based on the analytical results in Eqs.~\eqref{fractalCOEq} and~\eqref{fractalCUEq} using dashed black and red lines, respectively. }
  \label{GOE-GUE-Dq}
  \end{figure*}

Next, we examine the $N$-dependent multifractal dimensions and plot $\widetilde{D}_q^{(N)}$ vs $q$ in Fig.~\ref{GOE-GUE-Dq}, panel (a), using the same datasets as used for Fig.~\ref{GOE-GUE-MM-eigvec}. The data points (disks) from the simulation have been obtained by considering both spectral and ensemble averages and compared with solid colored lines obtained using Eqs.~\eqref{Dqtilcr} and \eqref{D1til} for various $\epsilon=\alpha^2 N$ values. In panel (b) we show the plots of $\widetilde{S}_q^{(\infty)}$ given by Eq.~\eqref{Sqinf} using solid colored lines. The disks show the data point corresponding to $\widetilde{S}_q^{(N)}$, obtained from the simulation as in panel (a). Also, In both the panels, the analytical results for the extreme cases of OE and UE have been shown using dashed black and red curves, respectively. We see good agreement between the analytical and simulation results, especially for lower $q$ values.  

\begin{figure*}[!tbp]
\centering
\includegraphics[width=0.9\textwidth]{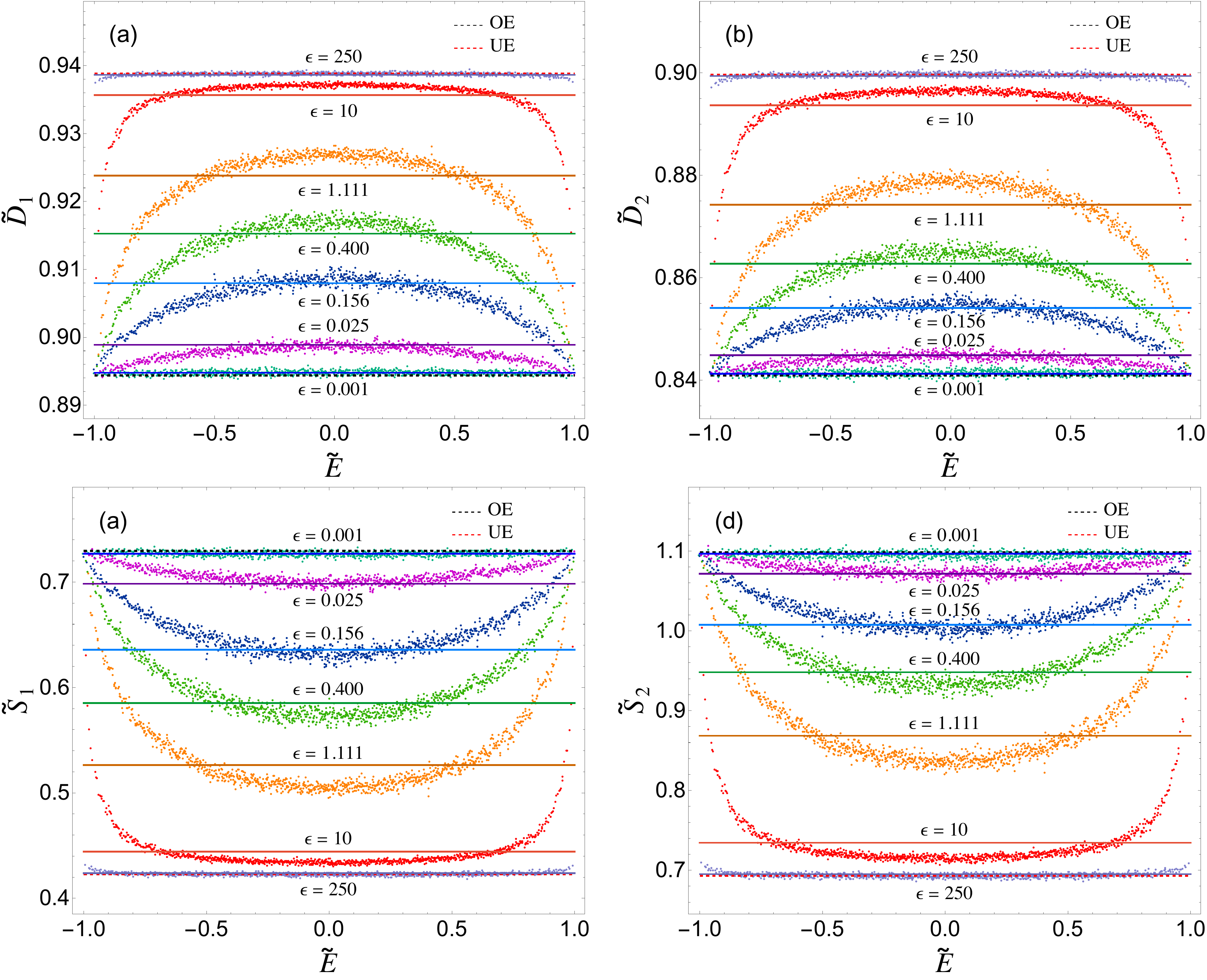}
\caption{Plots of ensemble averaged multifractal dimensions (a) $\tilde{D}_{1}$, (b) $\tilde{D}_{2}$ and (c) $\tilde{S}_{1}$, (d) $\tilde{S}_{2}$} for individual eigenvectors, sorted according to (ensemble averaged) eigenvalue magnitude. The numerical simulations for GOE-to-GUE crossover involves 200 matrices dimension of $N = 1000$. Analytical results are plotted with solid (crossover regime) and dashed lines (OE and UE extremes).
\label{MFD-MM-GOE-GUE}
\end{figure*}

In our next investigation, we again consider the GOE-to-GUE crossover matrix model, Eq.~\eqref{PM2}, and examine the quantities $\widetilde{D}_1^{(N)}, \widetilde{D}_2^{(N)}, \widetilde{S}_1^{(N)}, \widetilde{S}_2^{(N)}$ obtained from the individual eigenvectors and plot them with respect to the ensemble averaged ordered eigenvalues in Fig.~\ref{MFD-MM-GOE-GUE}. These are displayed as colored dots and are again based on 200 matrices of size $N=1000$. The solid horizontal lines in panels (a) and (b) correspond to analytical results for $\widetilde{D}_1^{(N)}$ and $\widetilde{D}_2^{(N)}$, whereas in panels (c) and (d) they correspond to $\widetilde{S}_1^{(\infty)}$ and $\widetilde{S}_2^{(\infty)}$. We notice that for the extreme cases (OE and UE), the behavior of eigenvectors is same throughout the spectrum. However, in the crossover regime the rate of symmetry crossover in the Gaussian ensemble depends on the location in the spectrum, owing to a nonuniform density of states for the initial ensemble (GOE)~\cite{PM1983,MP1983,KP2011,SKK2020}. The bulk eigenvalues undergo transition faster than the edge ones (those lying towards $\pm 1$ presently). Correspondingly, the associated eigenvectors also exhibit similar behavior in the crossover regime. In this case, the horizontal lines obtained using Eqs.~\eqref{Dqtilcr}, \eqref{D1til}, \eqref{Sqinf} and \eqref{S1inf} capture the average behavior of all the eigenvectors over the spectrum, which is also evident from the results in Fig.~\ref{GOE-GUE-Dq}. 

It is to be noted that with increasing $N$, the multifractal dimensions $\widetilde{D}_q^{(N)}$ for both OE and UE approach 1 and the difference between them tends to decrease. The corresponding values for the crossover ensemble, therefore, also gets squeezed between the former two and tend to 1. This convergence rate however, as already emphasized, is logarithmically slow. On the other hand, as $N$ increases, the values of $\widetilde{S}_q^{(N)}$ for distinct $q$ approach unique limits.

\section{Applications to quantum chaotic and many-body systems}
\label{QCMBS}

In this section we compute multifractal dimensions $\widetilde{D}_{1}$, $\widetilde{D}_{2}$ and the related quantities $\widetilde{S}_{1}$, $\widetilde{S}_{2}$ using simulations of two quantum chaotic systems, viz., the quantum kicked rotor (QKR) and a quarter Sinai billiard, and a many-body correlated spin chain model. Crossover from OE to UE class is realized in these systems by varying the respective relevant physical parameters. The results obtained are then contrasted with the RMT predictions of Secs.~\ref{MFoeue} and~\ref{MFouce}.

\subsection{Quantum kicked rotor}
\label{numre-QKR}

The quantum kicked rotor (QKR) is one of the foremost models for studying quantum chaos~\cite{Haake2010,SPK2022}. It has been exhaustively studied in several contexts and particularly in the verification of RMT analytical results in quantum chaotic systems~\cite{Haake2010}. The associated quantum dynamics of the kicked rotor may be described by the discrete time evolution operator (Floquet operator). A finite-dimensional model of QKR is realized by imposing a periodic boundary condition on the momentum basis which renders the corresponding classical phase space restricted to a two-dimensional torus~\cite{Casati1990,Izrailev1986,Izrailev1990}. The finite-dimensional Floquet matrix in the position basis is then given by
\begin{align}
\mathcal{U}_{mn} &= \frac{1}{N} \exp\left[-i \alpha \cos\left(\frac{2 \pi m}{N}+\theta_{0}\right)\right] \\ \nonumber \times & \sum_{l=-N'}^{N'}\exp\left[-i\left(\frac{l^2}{2}-\gamma l- \frac{2\pi l(m-n)}{N}\right)\right].
\end{align} 
In the above summation, $N'$ equals $ (N-1)/2 $ with $N$ odd and $m,n$ take the values $-N',-N'+1,...,N'-1,N'$. For the spectral fluctuation properties of the QKR to correspond to classical RMT ensembles very high values of the stochasticity parameter $\alpha$ is required~\cite{Casati1990, Izrailev1986, Izrailev1990,PRS1993,SP1997}. For studying time-reversal symmetry breaking in this system, we set $\theta_0=\pi/(2N)$ which corresponds to the fully broken parity symmetry case and then vary $\gamma$ starting from 0. This leads the eigenangle-spectrum of the matrix $\mathcal{U}$ to exhibit a transition from Circular Orthogonal Ensemble (COE) to Circular Unitary Ensemble (CUE) symmetry class~\cite{PS1991,PRS1993,SP1997,SKK2020}. 

\begin{figure*}[!tbp]
\centering
\includegraphics[width=17cm]{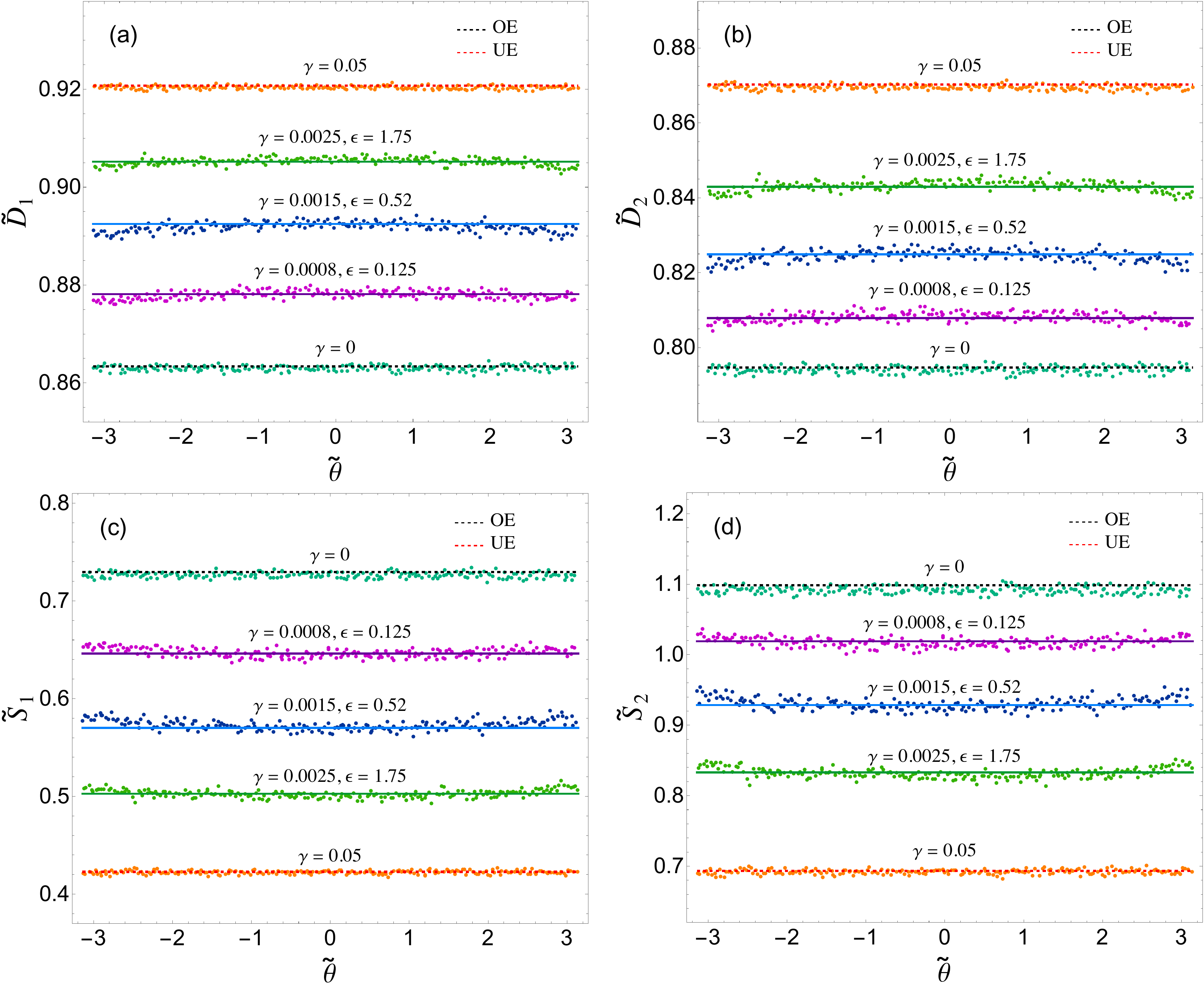}
\caption{Plots of (a) $\tilde{D}_{1}$, (b) $\tilde{D}_{2}$, (c) $\tilde{S}_{1}$ and (d) $\tilde{S}_{2}$ for QKR Floquet operator matrix of size $N =201$, averaged over an ensemble of 500 matrices, are plotted according to the magnitude of eigen-angles. The presentation scheme is same as in Fig.~\ref{MFD-MM-GOE-GUE}.}
\label{D1D2QKR}
\end{figure*}

Here we focus on the individual eigenvectors in the crossover region and calculate their fractal dimensions $\widetilde{D}_{1}$, $\widetilde{D}_{2}$, and the shifted and scaled quantities $\widetilde{S}_{1}$, $\widetilde{S}_{2}$, which are then plotted according to the magnitude of the ensemble-averaged eigenangles. We show the results for 500 matrices of dimension $N = 201$ in Fig.~\ref{D1D2QKR}. The ensemble is created by varying $\alpha$ in the vicinity of 20000. The colored dots correspond to the empirical results from the QKR simulations while analytical results are plotted with dashed or solid lines. We find that for $\gamma=0$ and  $\gamma=0.05$, the finite-dimension multifractal dimensions for the QKR are very close to the OE and UE values predicted by RMT. Moreover, the intermediate cases are fitted well using the crossover formulas by tuning the $\epsilon$ value. We note that here, unlike the case of Gaussian ensembles, the $\widetilde{D}_{1}^{(N)}$, $\widetilde{D}_{2}^{(N)}$, $\widetilde{S}_{1}^{(N)}$ and $\widetilde{S}_{2}^{(N)}$ do not show any pronounced dependence on location in the eigenangle-spectrum due to uniform rate of crossover for the eigenangles~\cite{PS1991,KP2011,Shukla1996,PRS1993,SP1997}.

\subsection{Quantum chaotic billiard}

Billiards serve as an important class of systems to study a variety of dynamical behaviour ranging from integrable (regular) motion to strongly chaotic behaviour. In the corresponding quantum domain one essentially deals with a \emph{particle in a box problem}, wherein the spectral fluctuations encode crucial information about the integrable or chaotic nature of the system. Due to Berry and Tabor, we know that the spectral fluctuation statistics of billiards which are classically integrable corresponds to Poissonian statistics~\cite{BT1977}. On the other hand, for billiards which are classically chaotic, the Bohigas-Giannoni-Schmit (BGS) conjecture asserts that the statistics of level fluctuations can be well explained by random matrix ensemble of a suitable symmetry class~\cite{BGS1984}. The classical dynamics in the Sinai billiard is well-known to be chaotic~\cite{S1963,S1970} and the corresponding quantum counterpart exhibits spectral fluctuations consistent with RMT. In our exploration, we have simulated a quarter Sinai billiard in KWANT Python package. It is a powerful open-source package intended for performing numerical calculations on tight-binding models~\cite{GA2014}. It should be noted that the discretized lattice version of billiard systems serve as approximations to the conitnuum case, however they also exhibit RMT based Wigner-Dyson spectral fluctuations~\cite{CLV1996,GRSZ2000,HLG2011,YLXHDGL2016,RRS2020}.

\begin{figure}[!tbp]
    \centering
    \includegraphics[width=0.6\linewidth]{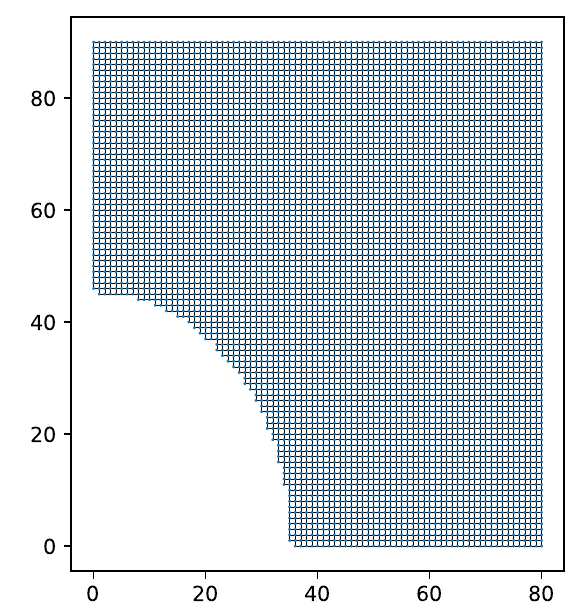}
    \caption{Representation of a quarter Sinai billiard simulated in KWANT, with the blue dots representing the lattice points. The billiard is constructed out of a rectangle of size $80\times 90$ with a quarter ellipse (semi-major axis length 45 and semi-minor axis length 35) carved out.}
    \label{Sinai}
\end{figure}

\begin{figure}[!tbp]
    \centering
    \includegraphics[width=0.9\linewidth]{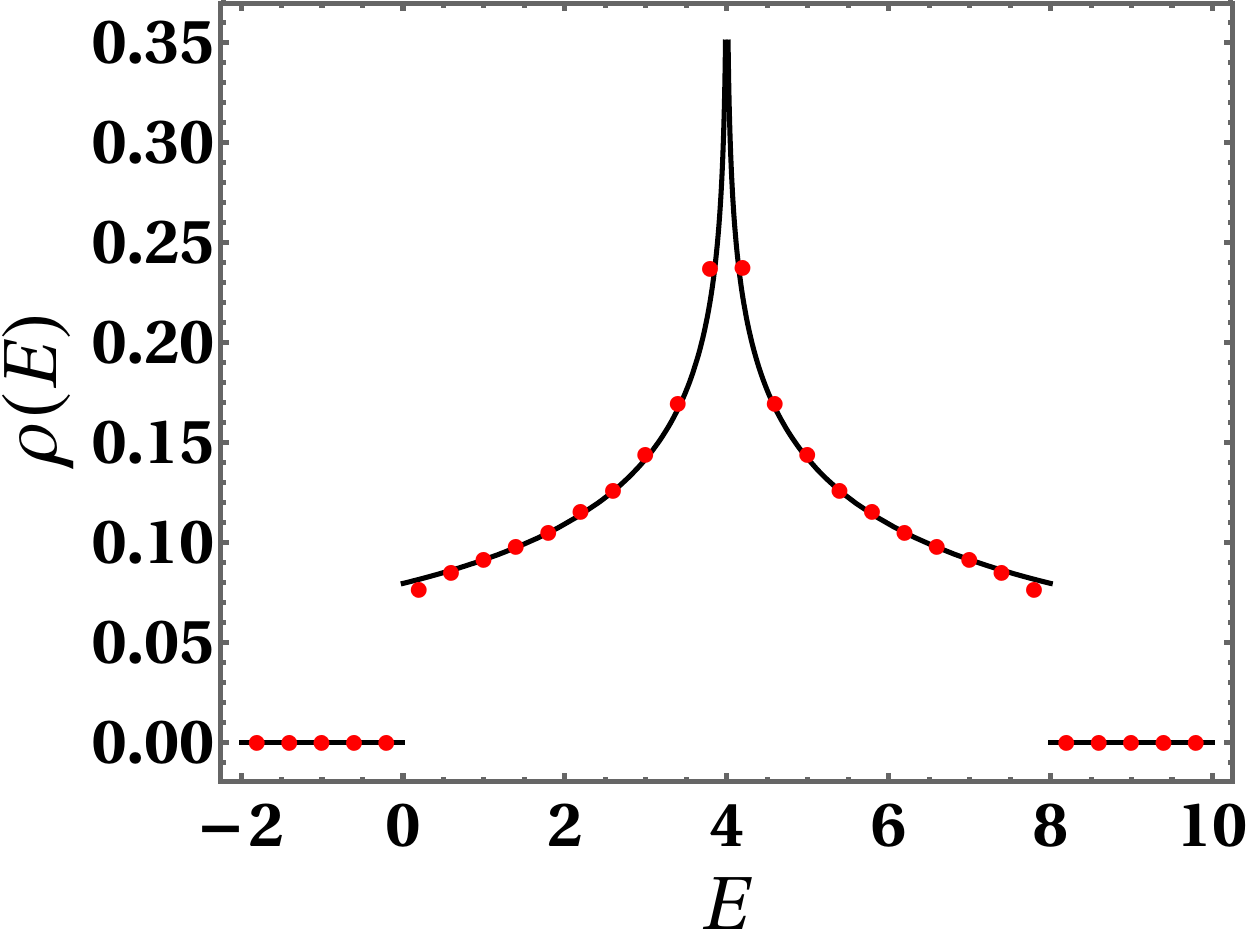}
    \caption{Density of states for the simulated quarter Sinai billiard, shown using red dots. The solid black line is analytical result based on Eq. \eqref{DOS}. The van Hove singularity at $E=4$ is clearly seen due to the system's lattice structure.}
    \label{DOS}
\end{figure}

We simulate the Sinai billiard by implementing the procedure followed in Ref. \cite{RRS2020}. The schematic is shown in Fig.~\ref{Sinai} along with the geometry details in the caption. The resulting Hamiltonian matrix is of dimension $N=6096$, which is large enough to render the computed the density of states (DOS) in good agreement with the well-known expression for DOS of a two-dimensional periodic square lattice~\cite{Economou2006},
\begin{equation}
\label{DOSeq}
    \rho(E)=\frac{1}{2\pi^{2}}\,\mathcal{K}\Bigg(\sqrt{1-\Big(\frac{E-4}{4}\Big)^{2}}\Bigg)\Theta(4-|E-4|);
\end{equation}
see Fig.~\ref{DOS}. In the equation above, $\mathcal{K}(u)$ is the complete elliptic integral of the first kind and $\Theta(u)$ is the Heaviside theta function. The density of states exhibits the van Hove singularity at $E=4$ which coincides with the onsite energy of our system, as expected from a lattice structure. Also, the DOS plot is symmetric about $E=4$.

\begin{figure*}[!tbp]
    \centering
    \includegraphics[width=18cm]{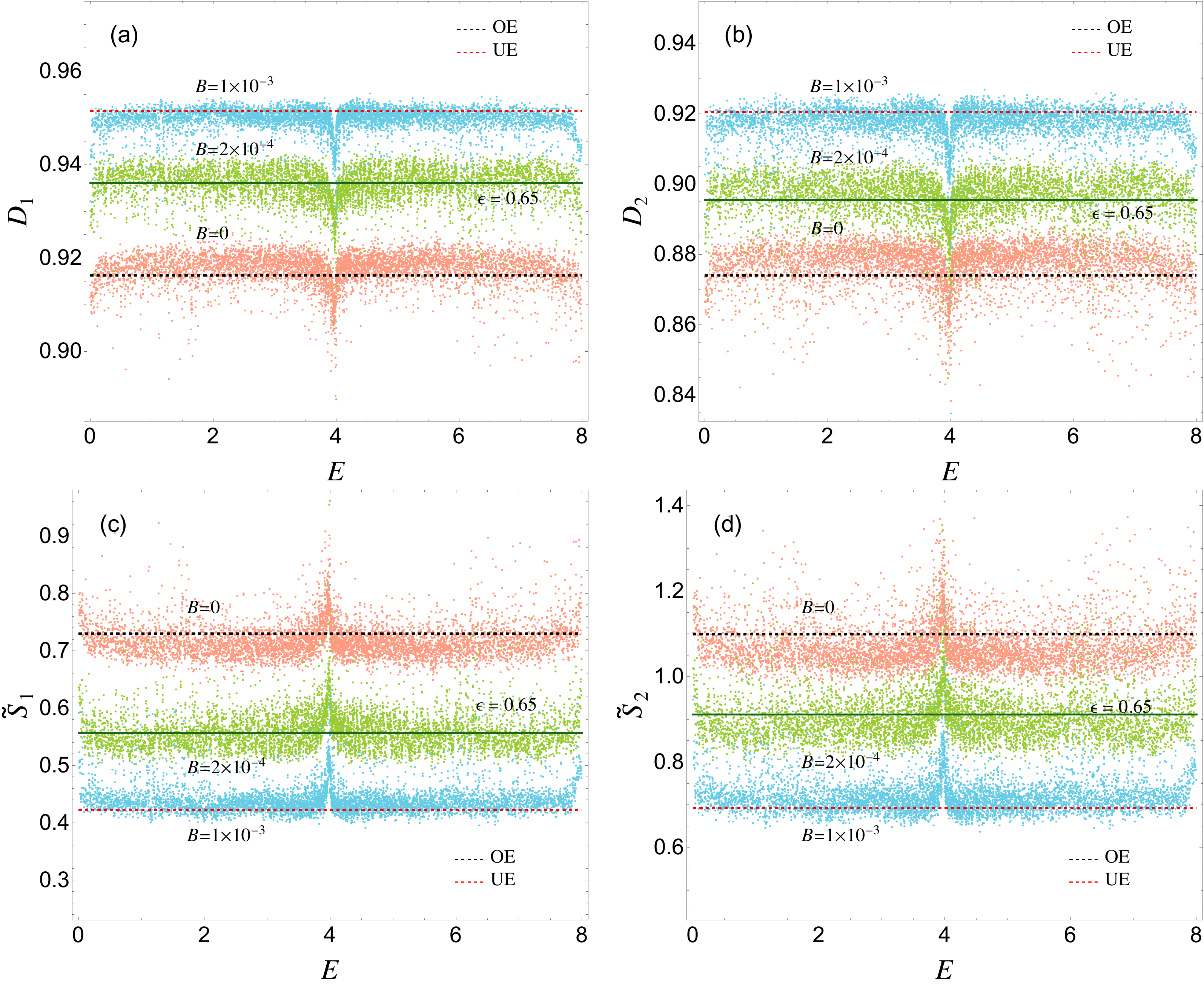}
    \caption{Plots of (a) $D_{1}$, (b) $D_{2}$ (c) $S_{1}$ and (d) $S_{2}$, shown using colored dots for individual eigenvectors of the quarter Sinai billiard Hamiltonian, sorted according to the energy eigenvalues. Three choices of magnetic field $B$ has been considered. In the bulk region, away from the edges and the van Hove singularity, the $B=0$ and $B=1\times10^{-3}$ results are close to OE and UE analytical evaluations (dashed black and red lines), while the $B=2\times 10^{-4}$ is found to be close to the crossover formula with $\epsilon=0.65$ (solid dark-green line).}
    \label{D1D2}
\end{figure*}

To study the orthogonal-to-unitary crossover in this system, we set up a magnetic field $B$ perpendicular to the plane of the system using the Peierls substitution method~\cite{Peierls1933,GA2014}. At $B=0$ the spectral fluctuations statistics corresponds to that of the orthogonal ensemble and on increasing the magnetic field, a gradually transition to the unitary ensemble statistics is observed. We sort the eigenvectors of the Hamiltonian according to the magnitude of the eigenvalues and compute $D_{1}^{(N)}$, $D_{2}^{(N)}$ and $S_{1}^{(N)}$, $S_{2}^{(N)}$ for each by considering three choices of the magnetic field, viz. $B=0$, $2\times 10^{-4}$, and $1\times10^{-3}$ in scaled units~\cite{RRS2020}. These results are plotted in Fig.~\ref{D1D2}. It should be noted that no ensemble averaging has been performed for the Sinai billiard results, whereas the RMT results are based on ensemble average. In the bulk region, away from the edges and that of the van Hove singularity, good agreement is observed with the OE and UE random matrix results for $B=0$ and $B=1\times10^{-3}$, respectively. Moreover, the intermediate case of $B=2\times 10^{-4}$ is fitted well using the crossover formula with $\epsilon=0.65$. Similarly other intermediate cases can also be shown, however, for clarity in the figure, we have considered only one intermediate situation. We also notice that for this system, compared to the QKR of the previous subsection and spin-chain model below, the values of the fractal dimensions are closer to 1 (ergodic behavior) due to the large $N$. Nonetheless it is possible to distinguish between the OE and UE symmetry classes and the crossover regime due to finiteness of $N$. Additionally, the quantities $S_{1}^{(N)}$, $S_{2}^{(N)}$ obtained from the billiard data acquire distinct values, close to the corresponding $N\to\infty$ limits which have been displayed using solid or dashed horizontal lines.


\subsection{Correlated spin chain systems}

We now examine a quantum many-body system, namely, a one-dimensional spin-1/2 correlated spin system and examine the corresponding multifractal dimensions. The Hamiltonian of this spin chain is given by~\cite{ARB2002,MM2014,KKSG2022},
\begin{equation}
    \label{Corrspin}
    \mathcal{H}= \sum\limits_{j = 1}^{L}[J \mathbf{S}_{j}.\mathbf{S}_{j+1}+h_{j}S^{z}_{j}+K\mathbf{S}_{j}.(\mathbf{S}_{j+1}\times \mathbf{S}_{j+2})].
\end{equation}
Here, $\mathbf{S}_{j}$ represents the spin operator at site $j$ and $\mathrm{S}^{z}_{j}$ gives the corresponding $z$ component. We implement periodic boundary condition, so that $\mathbf{S}_{L+k} = \mathbf{S}_{k}$. With the summation over sites considered, the first term in the above equation is the isotropic Heisenberg contribution, the second term represents the coupling of the spin system to a random magnetic field, and the third term gives the three-site scalar spin-chirality contribution. The parameter $J$ gives the nearest-neighbor exchange interaction, $h_{j}$ is the strength of site-dependent (zero-mean, $h^{2}$-variance random Gaussian number) magnetic field along the $z$ direction, and $K$ is the coupling constant for the spin-chirality term. 

\begin{figure*}[!tbp]
\centering
\includegraphics[width=17cm]{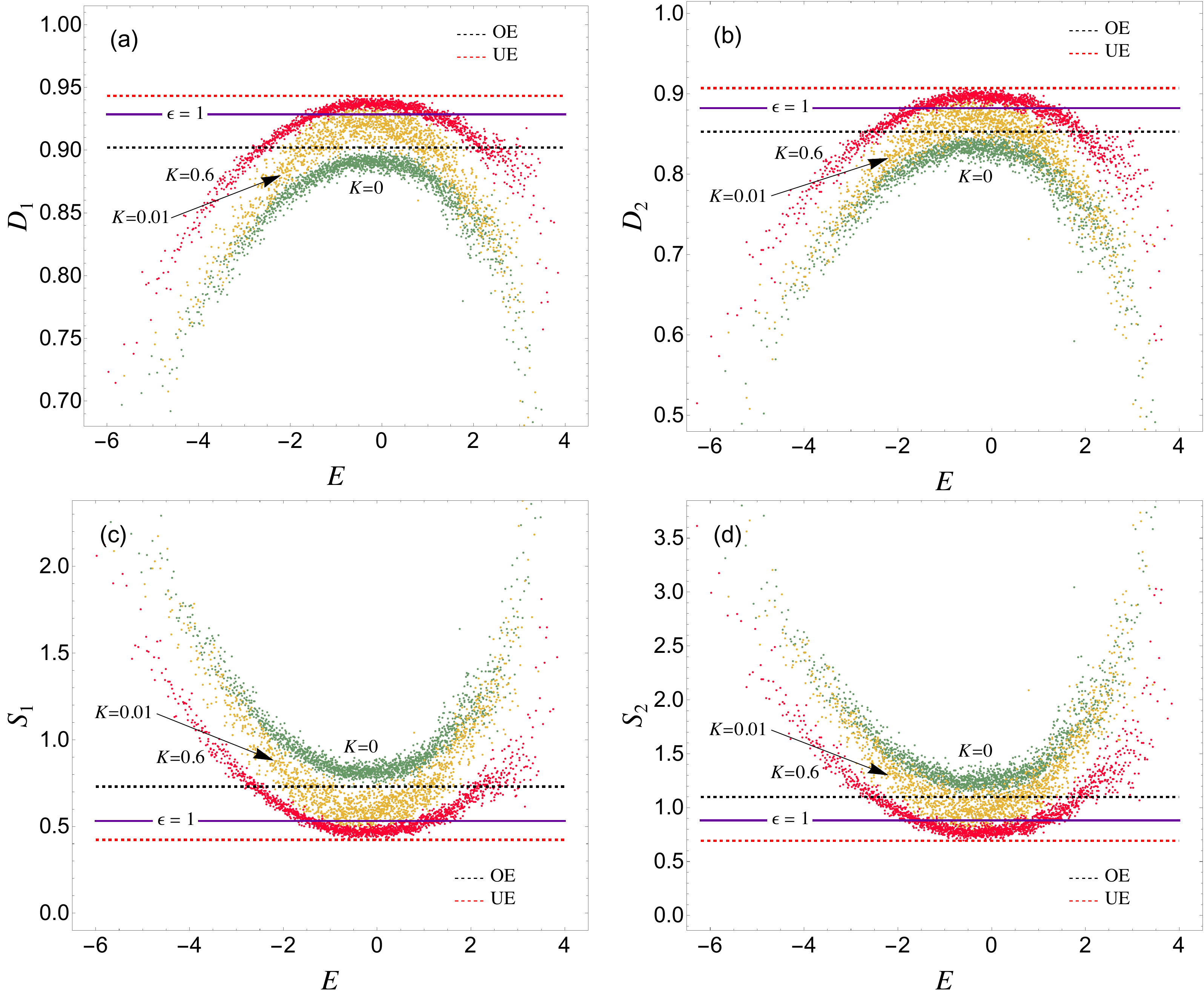}
\caption{Plots of (a) $D_{1}$, (b) $D_{2}$, (c) $S_{1}$ and (d) $S_{2}$, depicted using colored dots for individual eigenvectors of the spin chain Hamiltonian in Eq.~\eqref{Corrspin}, sorted according to the energy eigenvalues. Orthogonal to unitary crossover is realized by varying the spin chirality term controlling parameter $K$. Analytical OE and UE results (dashed black and red lines), along with the RMT crossover result for $\epsilon=1$ (solid purple line), are shown for comparison.}
\label{D1D2-N1716-SC} 
\end{figure*}

Various spectral-fluctuation crossovers in the above Hamiltonian can be achieved by tuning the strengths of the system parameters~\cite{ARB2002,MM2014,KKSG2022}. In particular, orthogonal-to-unitary crossover can be realized by increasing the value of $K$, the other parameters being fixed. One important feature of the above model is that the total $z$-component of spin, $\mathrm{S}^{z} = \sum_{j = 1}^{L} \mathrm{S}^{z}_{j}$ commutes with the Hamiltonian $\mathcal{H}$ and therefore when written in the corresponding basis, $\mathcal{H}$ acquires a block-diagonal form. For our symmetry crossover analysis, one such block has to be considered so as to avoid superposition of the independent subspectra associated with the individual blocks. For our calculations we consider a chain length of $L = 13$ for which the overall Hamiltonian size is $2^{13}$ = 8192. For $L = 13$, the largest diagonal block corresponds to $\mathrm{S}^{z}$ = 1/2 sector and is of size 1716. The parameter $K$ is varied from 0 to 0.6 to realize the symmetry crossover, with the other two parameters fixed at $J = 1$ and $h = 0.2$. We compute fractal dimensions $D_1^{(N)}$ and $D_2^{(N)}$ for the individual eigenvectors for this single block and show them with the sorted eigenvalues in Fig. \ref{D1D2-N1716-SC}. We observe very strong dependence of the multifractal dimensions on the location in the spectrum. The midspectrum values for $K=0$ and $0.6$ are close to random matrix OE and UE results, though noticeable deviations are visible. This is in agreement with a similar observation in Ref.~\cite{BHK2019} for OE case where the authors refer to mid-spectrum states as weakly ergodic. On the other hand, the states belonging to the spectrum edges exhibit multifractal behavior since the corresponding multifractal dimensions do not show any trend of approaching the value 1. The value $K=0.01$ in our analysis gives an intermediate case between OE and UE, which exhibits a similar dependence on the spectrum location. Moreover, in this case the random matrix crossover result with $\epsilon=1$ is found to be close to the multifractal dimension values in the midspectrum region.The shifted and scaled quantities $S_1^{(N)}$ and $S_2^{(N)}$ also reinforce these findings. Other $K$ values between 0 and $0.6$ also lead to intermediate statistics and can be fitted using the RMT formula with adequate $\epsilon$, however, for the sake of clarity in the figure, we have shown only one intermediate case. 

\section{Summary and Outlook}
\label{summ}

The eigenvectors of invariant random matrix ensembles serve as standard models for ergodic states, making their associated measures widely used for examining the localization properties of states in complex quantum systems. In the limit $N\to \infty$, all multifractal dimensions associated with the invariant random matrix ensembles converge to the value of 1. However, for large but finite $N$, significant finite-size effects are observed. These effects are strong enough to distinguish between different symmetry classes, such as orthogonal and unitary ones. Based on this crucial observation, it is reasonable to expect that the crossover regime between various invariant classes can also be quantified using finite-size statistics. In this work, we demonstrated that this expectation indeed holds true by providing semi-analytical expressions for the ensemble-averaged finite-dimensional multifractal dimensions in the orthogonal to unitary crossover ensemble. We validated these analytical results through Monte Carlo simulations of the underlying crossover random matrix model. Subsequently, we applied these expressions to investigate the quantum kicked rotor, a quarter Sinai billiard, and a correlated spin chain model. The symmetry crossover in these systems is achieved by varying the respective relevant system parameters. Additionally, we proposed shifted and scaled variants of the multifractal dimensions which acquire distinct values in the $N\to\infty$ limit and therefore may serve as better measures for distinguishing orthogonal and unitary symmetry classes and the crossover between them.

One interesting finding in our work concerns the dependence of multifractal dimension values on the location in the spectrum, within the Gaussian crossover ensemble. This dependence arises due to different transition rates in the spectrum caused by a nonuniform density of states. The results derived from a generic eigenvector component captures only the average behavior over the spectrum and fails to discern this location-dependence feature. Therefore, it becomes intriguing to examine the behavior of bulk and edge eigenvectors individually in the crossover regime and their impact on the resulting multifractal dimensions. Furthermore, such finite-dimensional analysis is also desirable for random matrix ensembles that remain unexplored from the perspective of multifractality.

\section{Acknowledgements}

A.S. is indebted to DST-INSPIRE for providing research fellowship (IF170612). She also thanks Debojyoti Kundu for fruitful discussions. S.K. acknowledges financial support provided by SERB, DST, Government of India, via Project No. CRG/2022/001751. 

\end{document}